\documentclass[conference, compsoc]{IEEEtran}
\ifCLASSOPTIONcompsoc
  \usepackage[nocompress]{cite}
\else
  \usepackage{cite}
\fi

\newif\ifanon
\anonfalse

\IEEEoverridecommandlockouts
\usepackage{amsmath,amssymb,amsfonts}
\usepackage{algorithmic}
\usepackage{graphicx}
\usepackage{textcomp}
\usepackage{xcolor}
\def\BibTeX{{\rm B\kern-.05em{\sc i\kern-.025em b}\kern-.08em
    T\kern-.1667em\lower.7ex\hbox{E}\kern-.125emX}}

\usepackage{dsfont}
\usepackage[nolist,nohyperlinks]{acronym}
\usepackage{soul}
\usepackage{tcolorbox}
\usepackage{tabularray}
\UseTblrLibrary{diagbox}
\usepackage{hyphenat}
\usepackage{hyperref}
\usepackage[libertine]{newtxmath}

\acrodef{IMU}[IMU]{\emph{Inertial-Measurement-Unit}}
\acrodef{SMPL}[SMPL]{\emph{Skinned Multi-Person Linear Model}}
\acrodef{VAE}[VAE]{\emph{Variational Autoencoder}}
\acrodef{AE}[AE]{\emph{Autoencoder}}
\acrodef{KL}[KL]{\emph{Kullback-Leibler divergence}}
\acrodef{CVAE}[CVAE]{\emph{Conditional Variational Autoencoder}}
\acrodef{MR}[MR]{\emph{Mixed Reality}}
\acrodef{VR}[VR]{\emph{Virtual Reality}}
\acrodef{DMM}[DMM]{\emph{Deep Motion Masking}}
\acrodef{CNN}[CNN]{\emph{Convolutional Neural Network}}
\acrodef{MSE}[MSE]{\emph{Mean Squared Error}}
\acrodef{PCA}[PCA]{\emph{principal component analysis}}
\acrodef{SVM}[SVM]{\emph{Support Vector Machine}}
\acrodef{RBF}[RBF]{\emph{radial basis function}}
\acrodef{ADE}[ADE]{\emph{average displacement error}}
\acrodef{FDE}[FDE]{\emph{final displacement error}}
\acrodef{ACC}[ACC]{\emph{accuracy}}

\newcommand{\horst}{Horst-DB}

\newif\ifcomments
\commentstrue
\ifcomments
\newcommand{\ts}[1]{\textcolor{blue}{\textbf{TS:} #1}}
\newcommand{\jt}[1]{\textcolor{purple}{\textbf{JT:} #1}}
\newcommand{\sh}[1]{\textcolor{olive}{\textbf{SH:} #1}}

\newcommand{\old}[1]{\textcolor{red}{#1}}
\else
\newcommand{\ts}[1]{}
\newcommand{\jt}[1]{}
\newcommand{\sh}[1]{}

\newcommand{\old}[1]{}
\fi

\begin{document}
\title{Pantomime: Motion Data Anonymization using Foundation Motion Models
}

\ifanon

\else
\title{Pantomime: Motion Data Anonymization using Foundation Motion Models
}

\author{\IEEEauthorblockN{1\textsuperscript{st} Simon Hanisch}
\IEEEauthorblockA{\textit{Centre for Tactile Internet (CeTI)} \\
\textit{Technical University Dresden}\\
Dresden, Germany \\
simon.hanisch@tu-dresden.de}
\and
\IEEEauthorblockN{2\textsuperscript{nd} Julian Todt}
\IEEEauthorblockA{\textit{KASTEL Security Research Labs} \\
\textit{Karlsruhe Institute of Technology}\\
Karlsruhe, Germany \\
julian.todt@kit.edu}
\and
\IEEEauthorblockN{3\textsuperscript{rd} Thorsten Strufe}
\IEEEauthorblockA{\textit{KASTEL Security Research Labs} \\
\textit{Karlsruhe Institute of Technology}\\
Karlsruhe, Germany \\
thorsten.strufe@kit.edu}
}
\fi

\maketitle

\begin{abstract}

Human motion is a behavioral biometric trait that can be used to identify individuals and infer private attributes such as medical conditions. This poses a serious threat to privacy as motion extraction from video and motion capture are increasingly used for a variety of applications, including mixed reality, robotics, medicine, and the quantified self. In order to protect the privacy of the tracked individuals, anonymization techniques that preserve the utility of the data are required. However, anonymizing motion data is a challenging task because there are many dependencies in motion sequences (such as physiological constraints) that, if ignored, make the anonymized motion sequence appear unnatural. In this paper, we propose \emph{Pantomime}, a full-body anonymization technique for motion data, which uses foundation motion models to generate motion sequences that adhere to the dependencies in the data, thus keeping the utility of the anonymized data high. Our results show that Pantomime can maintain the naturalness of the motion sequences while reducing the identification accuracy to 10\%.

\end{abstract}


\section{Introduction}\label{sec:introduction}

Human motion data is a rich source of information with various applications in medicine, \ac{MR}, or robotics. As motion tracking systems become cheaper and easier to use, we are now seeing more widespread adoption of motion tracking in everyday life. Examples of this are full-body motion tracking \ac{IMU}-suits and similar \ac{IMU}-trackers for \ac{MR} applications. Another example is the hand motion tracking most \ac{MR} headsets can now perform using video cameras. The captured motion data is used for applications such as the animation of digital avatars in social \ac{MR} scenarios~\cite{ma2021pixel}, the remote control~\cite{10035484} and training~\cite{terlemez2014master} of robots, as input modalities for \ac{MR} devices in general~\cite{han2020megatrack}, or for the monitoring of rehabilitation efforts in medicine~\cite{prill2021systematic}.

While these applications demonstrate how valuable motion data can be, it also carries risks, as it is behavioral biometric data that contains inherently sensitive information about the recorded person. 
Identifying people through their gait (pattern of walking) has long been a research topic for surveillance systems as it can be performed even on low-quality video data~\cite{wan_survey_2018}. 
In recent years, 3D motion capture data has also been investigated for gait recognition purposes, and it has been shown that the identification of individuals is simple and robust~\cite{horst_explaining_2019, DBLP:journals/popets/HanischMHLS23}. Further, it has been shown that diseases like Parkinson's can be inferred from gait data~\cite{abdulhay2018gait}.

Therefore, to prevent the user identification from motion data, while preserving its utility, we require anonymization techniques for full-body motion data. However, as prior work~\cite{DBLP:journals/popets/HanischMHLS23} has shown motion data is difficult to anonymize while maintaining high utility, because it contains a large number of correlations between data points, which an anonymization must consider. Otherwise, it is possible to break the protection or even reconstruct the original data from the anonymized data~\cite{DBLP:journals/popets/TodtHS24, DBLP:journals/popets/HanischTPES24}.

Therefore, we propose \emph{Pantomime}, the first anonymization for full-body motion sequences that is robust against re-identification while enabling high utility. Pantomime uses foundation motion models to hide the identity of people captured in motion data, which makes it applicable to different motion data formats and eliminates the requirement to train on the data it should anonymize. The use of foundation motion models allows Pantomime to project the motion data into the motion space. Pantomime then anonymizes the motion data in the motion space by adding random noise to it before decoding the motion data back into its original space. %
The advantage of this approach is that by adding noise to the motion space, we create a new plausible motion that is similar to the original. By increasing the noise, the new motion can be further away from the original motion, and thus anonymization can be increased at the expense of motion utility, allowing us to configure the privacy-utility tradeoff of Pantomime.

Furthermore, using foundation motion models has the advantage that they are trained on general motion data and therefore generalize well to different motions~\cite{rempe2021humor}. Because of this, Pantomime does not require the enrollment of specific users or the tuning of the model to specific datasets, as in previous work~\cite{nair2024deep}. Furthermore, Pantomime is applied to the motion data time-step by time-step, making it flexible in its application to motion data (e.g., when animating an avatar in a \ac{MR} chat application).

We investigate the privacy-utility tradeoff of Pantomime by measuring the utility via naturalness and by comparing the similarity of the anonymized sequence to the original using a user study. To measure the privacy protection we measure the person identification accuracy with a state-of-the-art biometric recognition system~\cite{horst_explaining_2019}.

Furthermore, we investigate how much of the individual components (e.g., body shape, joint rotations, etc.) of the motion data contribute to person identification. To better understand which component requires anonymization. The contributions of this work are as follows:

\begin{itemize}
    \item We propose Pantomime, the first general technique for anonymizing motion data that does not require training on the dataset to be anonymized and can configure its privacy-utility tradeoff.
    \item We evaluate Pantomime's privacy protection using two full-body motion capture datasets.
    \item We conduct a user study to investigate the naturalness and action similarity of the anonymized motion sequences.
\end{itemize}

The paper is organized as follows. First, we introduce the background in Section~\ref{sec:background}. We then examine the relevant related work in Section~\ref{sec:related_work} before describing Pantomime in Section~\ref{sec:pantomime}. We then describe how we evaluated Pantomime in Section~\ref{sec:evaluation} and present the results in Section~\ref{sec:results}. The paper ends with a discussion in Section~\ref{sec:discussion} and a short conclusion in Section~\ref{sec:conclusion}.

\section{Terminology \& Background}\label{sec:background}

\begin{figure}[ht]
    \centering
    \includegraphics[width=0.9\linewidth]{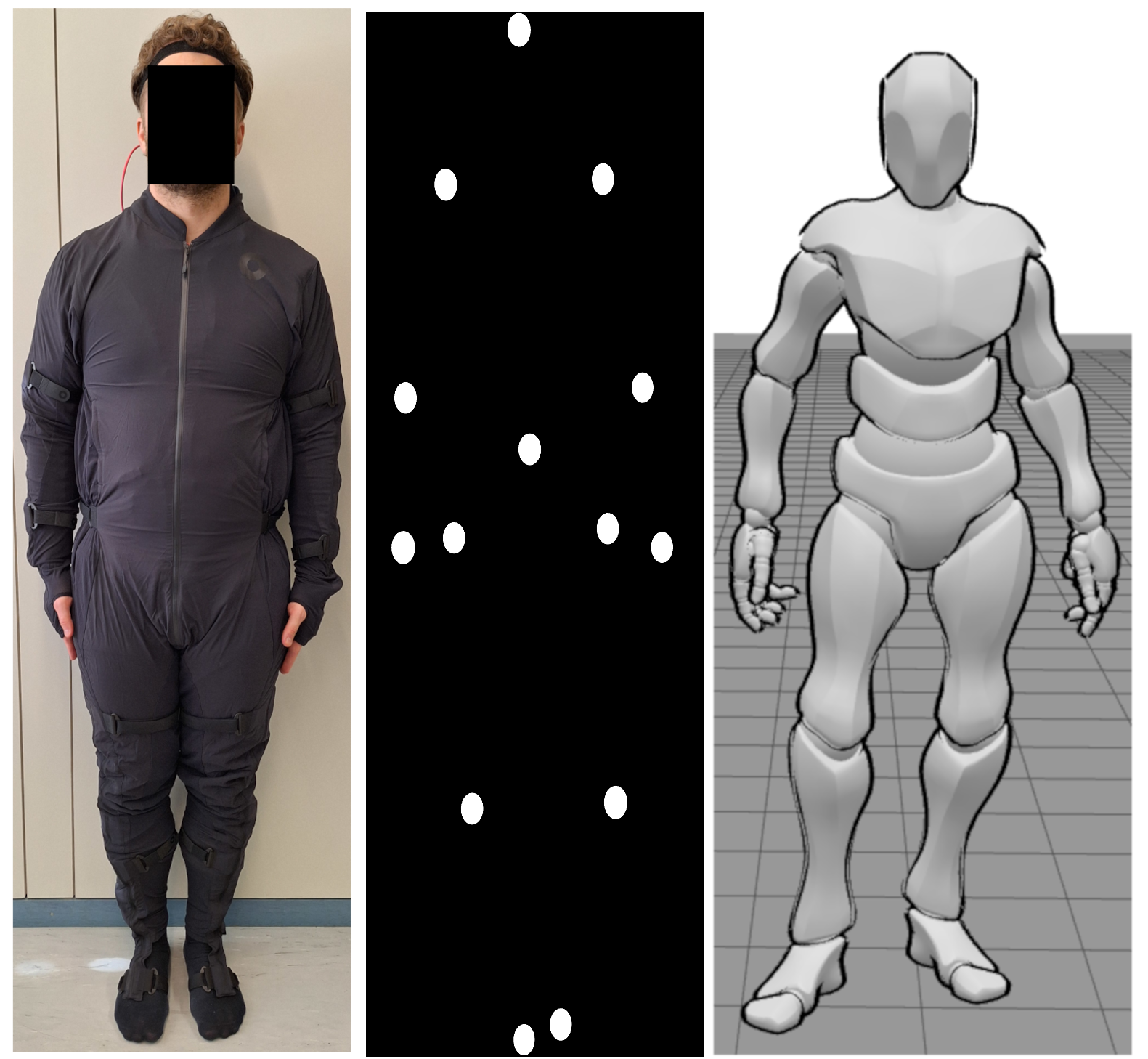}
    \caption{Left a person in an IMU-suit. Middle a point light display representation of motion data. Right a digital avatar.}\label{fig:point-light-display}
\end{figure}

In the following we introduce the terminology, autoencoder background, the SMPL body model, and foundation models used in this paper.

\subsection{Terminology}\label{sec:terminology}

Here we define the terminology for motion data and gait that we use in this paper.

A \emph{foundation model} is, as defined by Bomasan et al.~\cite{bommasani2021opportunities}: "... any model that is trained on broad data (generally using self-supervision at scale) that can be adapted (e.g., fine-tuned) to a wide range of downstream tasks". In this paper, we adapt motion models trained on a large motion corpus for motion sequence encoding and decoding to the task of anonymization.

Throughout this paper we will be working with \emph{motion sequences}. Motion sequences are time series of body poses. A \emph{body pose} consists of the position and orientation of all body parts of the captured person at a given point in time. In this paper we will use two representations of body poses. The first is the representation as a set of 3D points located at fixed body landmarks (e.g., shoulders) of the person. The second is the representation as a body shape plus the joint rotations of the individual body parts (e.g. arms, legs, ..).

Walking is a cyclic motion. For this work, we use the definition of a \emph{gait cycle} from Perry et al.'~\cite{perry2010gait}. There, a full gait cycle is completed when the feet are in the same relative position to each other as they were at the beginning of the cycle.

For our user study we use \emph{point-light displays}~\cite{deLussanet2018}. Point-light displays (see Figure~\ref{fig:point-light-display}) are a well-established method from psychology for representing motion data. Instead of displaying the whole body, only certain landmarks of the body are displayed as white dots on a black background. The advantage of point-light displays is that they allow viewers to judge the execution of a motion without being influenced by the appearance of the person performing the motion.

\subsection{Autoencoders}

Pantomime uses foundation models that use variants of \ac{AE} architectures, and relies on some of the features of these architectures. This is because anonymization is performed in their latent space. Here we focus on the features of \ac{AE} and its variants that are important for Pantomime.

An \ac{AE}~\cite{kramer1991nonlinear} is a machine learning architecture consisting of an encoder model and a decoder model. The encoder's goal is to translate the data into a much smaller latent space, which the decoder then translates back into the original data space. In other words, the encoder performs a compression of the data and the decoder decompresses it back to its original form. The overall goal of the model is to learn an efficient coding of the input data. Since the output of the model should be the same as the input, it can be trained unsupervised by using a reconstruction loss (e.g., \ac{MSE}) between the encoder input and the decoder output.
\ac{VAE}~\cite{kingma2022autoencodingvariationalbayes} is a type of \ac{AE} that, instead of learning a discrete latent code for a given input, maps the input to the parameters of a probability distribution.
A discrete latent code is then drawn from this probability distribution and decoded into the input space by the decoder. When training a \ac{VAE}, the shape of the learned latent distribution is often regularized by using a Kullback-Leibler divergence~\cite{10.1214/aoms/1177729694} to resemble a normal distribution. In this case, the learned latent space can be interpreted as a mixture of normal distributions.

Several \ac{VAE} variants exist. $\beta$-\ac{VAE}s~\cite{higgins2017betavae} use a weighted Kullback-Leibler divergence to increase its influence in the loss and thereby force a disentanglement of the dimensions of the latent code~\cite{sohn2015learning, burgess2018understanding}. Another type of \ac{VAE} is a \ac{CVAE}~\cite{sohn2015learning} that uses an additional label to constrain the latent code of an input to be deterministic. In this way, only the label can be fed to the decoder to generate a sample that belongs to the class of the input label.

\subsection{SMPL Body Model}

The \ac{SMPL}~\cite{SMPL:2015} model is a body shape model that uses blend shapes to represent the human body shape for different poses.
The \ac{SMPL} decomposes the body shape into a fixed identity-based body shape and a variable shape that depends on the body pose, which is represented as joint rotations.
Due to this decomposition into static and dynamic body pose, the SMPL model respects the body shape deformation that occurs in different poses, i.e. soft tissue deformations when a person is moving compared to when the person is standing still. The \ac{SMPL} model has been trained to minimize the reconstruction error of high-resolution 3D body scans.
For the remainder of the paper, the \ac{SMPL} model is defined as a differentiable function $M(r, \Phi, \Theta, \beta)$ mapping the root translation $r \in \mathds{R}^{3}$, root rotation $\Phi \in \mathds{R}^{3}$, body pose as joint angles $\Theta \in \mathds{R}^{3\times21}$, and identity-based body shape parameters $\beta \in \mathds{R}^{16}$ to the vertices $V \in \mathds{R}^{3x6890}$. The joint positions $J \in \mathds{R}^{3\times22}$ can be calculated from the vertices using a regressor matrix.

To find the SMPL model representation of a given body pose, we perform \emph{3D body fitting}. The goal of the fitting is to find the best parameters for the SMPL model to fit the given data of the body pose (e.g. all limb positions of a person). %

\subsection{Foundation Models}
In the following, we will introduce the two foundation motion models VPoser and HuMoR that we use for Pantomime.

\subsubsection{VPoser}
VPoser~\cite{SMPL-X:2019} is a \ac{VAE} that has learned the probability distribution of plausible human body poses. It takes a single body pose as input and tries to output the same body pose. Since it models plausible poses, it can be used to judge whether a given pose is plausible or not. This is used to fit the SMPL body model to given motion data by using the model as a loss.

VPoser uses the normal distribution $\mathcal{N}(\mu, \sigma)$ for its latent space.
The VPoser encoder takes body poses as joint angles $\Theta$ as defined by the SMPL body model and outputs a latent code $z$ given the learned parameters $\phiup$ and $\thetaup$. The encoder can be formulated as $q_\phiup(z_t|\Theta_t) = \mathcal{N}(z_t;\mu_\phiup(\Theta_t), \sigma_\phiup(\Theta_t))$. The latent code is then mapped back into the original pose space by the decoder $p_\thetaup(\Theta_t| z_t)$ to get the reconstructed pose $\hat{\Theta}_t$.

\subsubsection{HuMoR}
HuMoR~\cite{rempe2021humor} is a \ac{CVAE} that has learned the transition from one pose to the next. Unlike VPoser, it uses two successive poses as input to its encoder and then tries to output the transition from the first to the second pose with its decoder. With its focus on pose transitions, HuMoR effectively models the distribution of human motion. Like VPoser, HuMoR can be used to fit body models to motion sequences.

HuMoR represents the state of a moving person as a matrix $x \in \mathds{R}^{69}$ consisting of root translation, root orientation, body joint angles, body joint positions, and the velocities of root translation, root orientation, and joint positions. The encoder of \nohyphens{HuMoR} takes two sequential states $x_t$ and $x_{t-1}$ as input and generates a latent code $z_t$ via the parameterization of a normal distribution ${q_\phiup(z_t|x_t, x_{t-1}) = \mathcal{N}(z_t;\mu_\phiup(x_t, x_{t-1}), \sigma_\phiup(x_t, x_{t-1}))}$ with the learned parameters $\phiup$. The latent code $z_t$ is then decoded by the decoder $p_\thetaup(x_t| z_t, x_{t-1})$ into the state change $\Delta_\thetaup$ and the ground contacts $c_{t}$ with the learned parameters $\thetaup$. The reconstructed motion state $\hat{x}_t$ is then ${\hat{x}_t = x_{t-1} + \Delta_{\thetaup}(z_t, x_{t-1})}$.

\section{Related Work}\label{sec:related_work}

Below is an overview of research in the emerging field of motion anonymization that attempts to prevent identification. We categorize the works according to the type of motion data they anonymize.

\subsection{Body Shape anonymization}
Sattar et al.~\cite{sattar2020body} investigated the privacy of body shapes extracted from single images. They show that a person's body shape is considered private information by performing a small user study, and propose an adversarial perturbation to prevent the automatic extraction of shape information from images. While this work is not motion anonymization, it highlights the need for anonymizing the body shape information that is implicitly contained in motion data.

\subsection{Video Anonymizations}
After motion data was identified as private information, some early anonymization approaches were proposed to protect it from inference. An early approach to anonymize human gait in videos was proposed by Tieu et al.~\cite{tieu2017approach}, where the silhouette and a noise silhouette of a walking person are mixed using a \ac{CNN} trained similarly to a \ac{AE} with the weighted \ac{MSE} of the original and noise silhouettes as the reconstruction loss. Hirose et al.~\cite{hirose_anonymization_2022, hirose2019Anonymization} investigated how gait videos can be anonymized by deforming the silhouette of the walking person and then computing a new texture for it to preserve utility. The approach uses a \ac{AE} that learns a code for the silhouette of the walking person. It also estimates the phase of the gait. Both the phase and the shape coding are then anonymized before being fed to the decoder to generate a new gait silhouette. The anonymization of the shape code is done using a $k$-anonymity approach, where the $k$-nearest shape codes are mixed and averaged with the encoded one. The last step is to compute a new texture for the silhouette. Romero et al. propose GaitGuard~\cite{romero2024gaitguardprivategaitmixed}, a real-time anonymization for gait in the camera view of \ac{MR} devices, where the anonymization is offloaded to a local server. For the anonymization itself, different approaches such as applying Gaussian blur have been tested. The goal of GaitGuard is to prevent pose extraction.

The anonymizations described above only work for 2D gait videos and are therefore limited, as they cannot be applied to 3D motion captures and do not work for motion data in general. However, the use of a \ac{AE} for silhouette anonymization can be adapted for 3D motion data.

\subsection{Egocentric Video Anonymization}
Thapar and Arora~\cite{Thapar_2021_ICCV} propose an anonymization technique for egocentric gait videos. In egocentric videos, the person using the camera is never seen directly, as the camera captures the person's field of view. However, it is still possible to infer the identity of the person recording the video. The anonymization technique uses machine learning to estimate the rotation of the camera and then adds the rotations of another video to anonymize it. Egocentric video anonymization can be considered a special case of 3D motion anonymization, since only the 3D rotation of the camera is anonymized. Full body motion data consists of a set of 3D positions and rotations, so this approach is not applicable.

\subsection{Motion Capture Anonymization}
Malek-Pdjaski and Deligianni~\cite{malekpodjaski_towards_2021} developed an anonymization technique for 3D motion capture that extracts features that do not allow identification but can still be used for affect recognition.
They attempt to separate the information needed for affect recognition from the information used for identification by using two \ac{AE}. One \ac{AE} is trained to be subject-specific and one \ac{AE} is trained to be affect-specific.
The disadvantage of this approach is that the \ac{AE}s have to be trained on the dataset to be anonymized. %
Moon et al.~\cite{moon2023anonymization} proposed an adversarial anonymization scheme for 3D motion capture data in which a machine learning model is trained to minimize the identity recognition and maximize the action recognition.
Both approaches are not suitable for preserving the naturalness of the motion data (as determined by a user study), since the benefit must be quantifiable so that it can be used as a loss in the training of these approaches.

\paragraph{Simple Anonymizations}
Moore et al.~\cite{Moore2021Personal} suggest using only the velocities of the motion sequences and Miller et al.~\cite{miller2020personal} suggest using only the joint rotations to reduce identifiability. Meng et. al.~\cite{meng2024avatars} additional propose adding noise to the joint rotations.
Hanisch et al.~\cite{DBLP:journals/popets/HanischMHLS23} investigated the effect of various simple anonymization techniques such as noise injection or removal of specific body parts from 3D motion capture data. All of these simple anonymization techniques fail to achieve a good privacy-utility tradeoff and show that identity recognition is possible even on heavily corrupted data.

\paragraph{Mixed Reality Anonymizations}
Nair et al. investigate how the (motion) data collected by \ac{VR} headsets can be anonymized. They proposed a framework called MetaGuard~\cite{Nair2023Going} which claims to protect various attributes collected by \ac{VR} headsets. Their second proposal \ac{DMM}~\cite{nair2024deep} is a machine learning approach which reduces identity similarity while maintaining action similarity. The evaluation of the approaches performed in \cite{nair2024deep} shows that \ac{DMM} is effective and can anonymize motion sequences of different datasets against different attackers, but MetaGuard is not effective and does not provide sufficient protection. The drawback of \ac{DMM} is that it requires a large training dataset and is application specific for the data it was trained on. %

\subsection{Summary}
In summary, previous work is limited in several ways. They either require large amounts of training data, are action specific (gait only), cannot be applied to full body motion data, do not preserve the utility of the data, or are simply not effective. %

\section{Methodology}\label{sec:pantomime}

As previous work~\cite{DBLP:journals/popets/HanischMHLS23, nair2024deep} has shown, anonymizing motion data is a challenging task. The main problem is that motion data contains a large number of dependencies between the individual tracked points, and constraints such as the maximum degree of flexion of certain joints. In addition, the physiology of the person performing the motion is important, as it strongly influences how motions are executed to perform the same action. For example, a tall person will bend their shoulder joint differently to grab an object from a table than a shorter person in the same situation. Because of all these dependencies, directly modifying the data is either not effective because the dependencies can be used to reconstruct the original data, or the modification has to be very strong, which greatly reduces the utility of the motion data. Another interpretation of the dependencies is that they represent redundancy in the data, since the true dimension of the motion data, which can be changed independently, is much smaller than the recorded positions.

The main idea of Pantomime is to remove as many of the dependencies described above as possible before performing anonymization, and then reintroduce the dependencies after anonymization to generate a new sequence of motion data. We perform the dependency removal by mapping a motion sequence into the motion space of a foundation motion model. We then anonymize the motion sequence in the motion space by adding noise to it. Finally, we map the anonymized motion sequence back to its original position space.

\subsection{Requirements}
The two main goals of motion anonymization are to prevent the identification of an individual from their respective motion sequences, and to preserve the utility of the motion sequences for the application for which they are intended. For Pantomime, the utility goals are naturalness and action similarity. Naturalness means that the motion sequence appears as a genuine motion sequence to a human observer. And action similarity means that the action in the anonymized sequence should be as similar as possible to the original one. We chose naturalness because for many applications of full-body motion data, such as social interactions in \ac{MR}, it is important that the motions appear believable and realistic. The action similarity should prevent our anonymized motion sequences from deviating too much from the original ones, otherwise we could just generate random motion sequences to satisfy the naturalness goal.
In addition, it would be beneficial to meet the following requirements derived from common applications of motion data. The first application requirement is the applicability of anonymization to full-body tracking, since even from sparse input data, such as tracking from \ac{MR} devices, the full-body pose can be estimated. The second is that the anonymization should be general with respect to the format in which the motion data is captured, since motion capture systems vary in the number of points tracked and the specific body landmarks that are captured. %

\subsection{Adversary Model}

The main goal of the adversary is to identify which person belongs to which motion sequence. We assume a strong adversary as described by Hanisch et al.~\cite{DBLP:journals/popets/HanischMHLS23} in that the adversary is aware of the existing anonymization and can adapt his attack to it. The adversary becomes stronger by adapting its attack, resulting in more rigorous evaluation results. To achieve his goal, the adversary uses a biometric recognition system to perform identification. To train the recognition system and learn templates for each person under attack, the adversary has access to clear motion data of these people. Since the adversary knows the anonymization used and its parameters, they can anonymize the clear data. By training the recognition system on anonymized data, the attacker can adapt the recognition system to the changes that the anonymization makes to the original data, and thus adapt their identification attack to the anonymization, thus posing a greater challenge to the anonymization.

\subsection{Anonymizing Human Motion}

We will now explain how Pantomime works. As a first step, we unify different motion capture formats into the format of the SMPL model by performing a fitting step. This step is necessary because we want Pantomime to be general enough to work with different formats of motion data. After transforming the original data into SMPL data, we map the data into the motion space of a foundation motion model and then perform anonymization by adding noise to the data. The intuition here is that since the motion space encodes plausible motions, by modifying the input data in this space, we end up with a plausible motion for the output of the anonymization. In contrast, performing anonymization by adding noise to the original position data quickly leads to implausible motions because the individual points are modified without adhering to the given physiology of the body or physics. The final step is to decode the anonymized motion space data back into the SMPL model format and then back into the original motion data format.

\begin{figure*}[ht]
    \centering
    \includegraphics[width=0.95\textwidth]{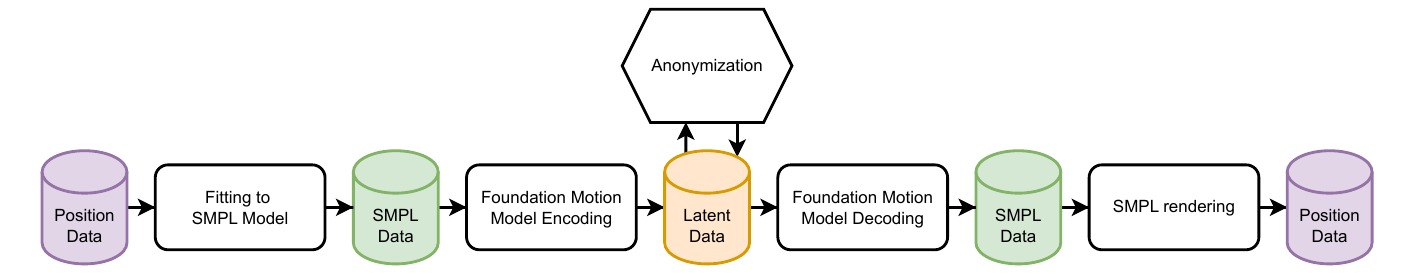}
    \caption{The Pantomime anonymization pipeline.}
    \label{fig:pantomime-pipeline}
\end{figure*}

In the following, we describe the different steps of our anonymization pipeline for Pantomime, an overview of the whole process can be seen in Figure~\ref{fig:pantomime-pipeline}.

\subsubsection{SMPL Model Fitting}\label{ssec:model_fitting}

To unify the motion data formats, we perform a fitting of our original motion sequence to the SMPL body model using the HuMoR~\cite{rempe2021humor} fitting method. This way all of our motion sequences have the same format and can be used with the same foundation motion models. The fitting process is a function optimization that is performed for each pose of a motion sequence to find the best SMPL parameters to represent that pose.

 Given an observed motion sequence, we try to find the parameters of the \ac{SMPL} body model $M(r, \Phi, \Theta, \beta)$ that best describe the sequence. To perform this function optimization, three loss terms must be minimized. The first is the plausibility of the motion. Here we use either VPoser or HuMoR as a prior to measure the plausibility of a given pose (VPoser) or the plausibility of a given pose transition (HuMoR). The second loss term is a reconstruction loss comparing the original with the found SMPL joint positions using \ac{MSE}. Since the skeletons between motion data representations may have different kinematic structures, we define a mapping between the SMPL body model skeleton and the dataset we are trying to fit to compute this reconstruction loss. The third loss is a regularization term that enforces consistency of bone length, ground contact, ground position, and body shape across the entire motion sequence (i.e., bone length and body shape should not change much during a single motion sequence). The weighting of the three loss parts is dataset specific and is determined by performing a hyperparameter optimization to find the optimal values. A more detailed description of the loss terms can be found in ~\cite{rempe2021humor}.

This first step can also be seen as a decoupling of the dependencies, since the original position data is split into the body shape $\beta$ and the joint angles $\Theta$, effectively decoupling these two aspects of the data.

\subsubsection{Encoding the Motion Sequences}

Next, we encode the motion sequences into the latent space of a foundation motion model. The foundation model used for this step is interchangeable, as Pantomime only requires that it be a \ac{VAE} that encodes from the SMPL parameters to a latent space of plausible poses or pose transitions and back to the SMPL parameters. The rationale behind this requirement is that \ac{VAE}s compress the original data by removing dependencies~\cite{sohn2015learning, burgess2018understanding} (i.e., correlations) between data points, and thus our anonymization no longer needs to adhere to these dependencies for effective anonymization. In other words, by performing anonymization in the space of plausible motions, the resulting anonymized motion is itself plausible, thus preserving the utility of the data. In this paper, we use VPoser and HuMoR as two possible foundation motion models that satisfy this requirement. However, it is important to note that VPoser only takes the body pose (joint angles $\Theta$) as input, while HuMoR also requires the root orientation, root translation, and body shape.

\subsubsection{Anonymizing the Latent Code}
Now we anonymize the latent code. To do this, we draw a noise vector $p$ from a normal distribution $\mathcal{N}(\mu, \sigma)$ with $\mu=0$ and $\sigma=1$ and add it to the latent code $z_t$ at time step $t$.
We chose normal distributed noise because \ac{VAE}s regularize their latent space to be normal distributed and adding two normal distributed random variables results in a normal distributed sum~\cite{lemons2002introduction}. Since the decoder expects a normally distributed random variable as input, this should result in the least utility loss. We chose the mean $\mu$ to be 0 and the $\sigma$ to be 1 because we do not want to introduce a bias into the anonymized motion sequences. With a mean of 0, the resulting distribution of anonymized motion sequences will cluster around the original motion sequence.
The noise vector $p$ is scaled by the scalar $\gamma$ to make the anonymization configurable by increasing the distance in motion space to the original sequence. There are two different modes for adding the noise. In \emph{variable} we draw a new noise vector $p_t$ for each time-step $z_t$, giving us $a_{t} = \gamma p_t + z_t$ for the anonymized latent code $a_t$, while in \emph{static} we add the same noise vector $p$ to each $z_t$, giving us $a_{t} = \gamma p + z_t$.

\subsubsection{Decoding the Motion Sequence}

The last step is to decode the anonymized latent code $a_t$ into the parameters of the SMPL body model using the motion model, and from there back into the original position format. Note that for Pantomime, we only focus on the anonymization of the body pose, the $\Theta$ parameter of the SMPL body model. We do not consider the anonymization of the body shape, the root translation, and the root orientation and remove them by setting their respective parameters to zero. We do this because we assume that a person who wants to be anonymous will choose a digital body that does not resemble them, including the body shape. The root translation and orientation can be estimated from the resulting motion sequence.

\section{Evaluation} \label{sec:evaluation}

We now evaluate the privacy-utility tradeoff of Pantomime to understand how much noise must be added for effective anonymization and how much utility is retained. We also test the assumptions we made when designing Pantomime.

\subsection{Datasets}

We select our datasets to include a large number of full body motion capture sequences with a preference for the gait task, as this has been shown to be highly identifiable. We specifically did not select the AMASS~\cite{AMASS:ICCV:2019} dataset, or any of the datasets included in it, because AMASS was used to train VPoser and HuMoR.

For our evaluation, we use the \textit{CeTI-Locomotion}~\cite{CeTILocomotion} and \horst~\cite{horst_explaining_2019} datasets. CeTI-Locomotion includes a variety of walking modalities as well as sit-to-stand exercises in which participants stand up and then sit down as quickly as possible. The dataset was recorded from 50 healthy participants using an \ac{IMU} suit that captures the relative motion of each body part. In combination with the anthropocentric measurements of the participants, the 17 body segment positions are calculated. Horst-DB contains only one walking modality and was recorded using optical motion tracking with 54 reflective markers attached to body landmarks such as joints or the iliac crest. The resulting data are scalar values for the x,y,z positions of each of the markers. The main difference between the two datasets is that CeTI-Locomotion includes several different walking modalities (normal, fast, carrying a backpack, and carrying a bottle crate) plus an additional sit-to-stand exercise. Also, IMU tracking is less accurate than optical marker tracking, which is considered the gold standard of motion tracking. See Table~\ref{tab:overview_datasets} for a comparison of the two datasets.

\begin{table}[!h]
    \caption{Overview numbers of the used evaluation dataset}\label{tab:overview_datasets}
    \centering
    \begin{tabular}{|l|l|l|l|l|}
    \hline
        Name & Points & Participants & Tasks & Samples \\ \hline
        CeTI-Locomotion & 17 & 50 & 5 & 4672\\ \hline
        \horst & 54 & 57 & 1 & 1140 \\ \hline
    \end{tabular}
\end{table}

\subsection{Implementation}

Here we describe the implementation details for both Pantomime and the biometric recognition systems we will use for our experiments.

\subsubsection{Data preparation}

We preprocess both of our evaluation datasets to have a frame rate of 30 Hz, the same as in the previous work~\cite{rempe2021humor}. For the Horst DB dataset, we additionally trim the samples to exactly one gait cycle (see Section~\ref{sec:terminology}) using the additional force plate data and a threshold to identify the first and last pose of the cycle, as described by Horst et al.~\cite{horst_explaining_2019}.

\subsubsection{SMPL parameter fitting}

Overall, the goal of the fitting is to find the parameters of the \ac{SMPL} model ($M(r, \Phi, \Theta, \beta)$) that match the positions of the input data as closely as possible, as well as to achieve a high plausibility with the used foundation motion model (here HuMoR or VPoser). The foundation motion model is used here to enforce the generation of only plausible motions for the \ac{SMPL} poses. To fit our two evaluation datasets we use the code of HuMoR~\cite{rempe2021humor}, which implements the whole process in three steps. In the first stage, the root translation and rotation are optimized using VPoser as a prior, in the second stage, the entire SMPL parameters are optimized using VPoser as a prior, and in the third stage, HuMoR is used as a prior for the optimization. Since the code performs the optimization first with VPoser and then with HuMoR, we use it to generate both fits.

To perform the fitting, we need a mapping from our motion data joint positions to the \ac{SMPL} joint positions (i.e., which positions in our data correspond to which position in the SMPL joints). We created the mapping manually as follows. For CeTI-Locomotion this mapping is undercomplete because CeTI-Locomotion has only 17 joints while the \ac{SMPL} model has 21. For the Horst DB, it is the other way around because the datasets tracked 54 points, some of which are ignored while others are combined to better match the joint positions of the \ac{SMPL} body model. Joint positions for which there is no matching position in the original data are set to infinity. To obtain good fitting results, we perform hyperparameter optimization on both datasets to find the weights for the fitting losses described in section~\ref{ssec:model_fitting}. Due to the length of the fitting process, we perform the hyperparameter optimization on 10 random motion sequences of each dataset and then use the found parameters for all of them.
For both datasets we had some motion sequences for which the SMPL fitting process failed (producing NaN values at some stage of the fitting process) and we were unable to obtain an SMPL body model representation, these sequences were excluded from the datasets. In both datasets this was less than 1\% of the total motion sequences.

As Pantomime removes the root translation from the sequences, we estimate a new root translation for the anonymized sequences by using the absolute trajectory of the right foot as the root translation.

\subsubsection{Biometric recognition system}

For the biometric recognition system, we adapt a state-of-the-art method used in previous work~\cite{DBLP:journals/popets/HanischMHLS23, horst_explaining_2019}, in which the motion sequence is resampled to a fixed 100 frames and then flattened into a single vector. The samples are then divided into 80\% training data and 20\% test data in a stratified fashion, while keeping the same percentage of samples per identity class. We use the same system for action recognition, but there we split the data so that a single participant's samples are either in the test data or in the training data. This should prevent the system from learning actions for specific people and help with generalization. The two datasets are then processed independently. Each dimension of the feature vector is min-max normalized before its dimensionality is reduced using a \ac{PCA}. As a classifier, we use a \ac{SVM} with a \ac{RBF} kernel, which is trained in a 5-fold stratified cross-validation procedure using balanced accuracy as a metric. The final result is obtained by running the \ac{SVM} on the test data. The recognition system is always trained on the anonymized data of the anonymization we are testing, as defined by our adversary model.

\subsubsection{Code Availability}

We implemented the biometric recognition system using Python, scikit-learn~\cite{scikit-learn} and PyTorch~\cite{NEURIPS2019_9015}. Pantomime itself is implemented in Python on top of the existing HuMoR code~\cite{rempe2021humor}. The anonymization and scoring code for Pantomime will be published with this paper.

\subsection{Experiments}

Here we detail the experiments we performed to investigate our underlying assumptions in designing Pantomime, and then to evaluate Pantomime's privacy-utility tradeoff.

\subsubsection{Assumptions \& Baseline} %

\paragraph{Baseline Identification \& Action Recognition}
First, we establish an identification and action recognition baseline on the original position data, against which we will later compare the anonymization results of Pantomime. This will allow us to evaluate how good Pantomime's privacy protection is and how much utility we lose as a result. In our baseline identification experiment \textbf{E1}, we train and test the biometric recognition system on the dataset to perform person identification. To do this, we split the dataset into a test and a training part of the dataset, with each person having different samples in both partitions. We then train the biometric recognition system on the training dataset in a supervised manner. We then determine the identification performance on the test set.
In the action recognition baseline experiment \textbf{E2}, we determine the baseline action recognition performance. We now train and test an action recognition system. Unlike for person identification, we split the dataset so that a person is either in the test or training dataset to avoid the system learning the unique action performance of a person and to better generalize. We then measure how well the system can identify the action performed in the sample.

\paragraph{Identification Potential SMPL Parameters}
The unification of the motion data in the SMPL body format splits the data into motion data (poses as joint angles~$\Theta$), body shape~$\beta$, root translation~$r$, and root orientation~$\Phi$. Pantomime focuses only on anonymizing the motion data of the poses. For the remaining SMPL parameters, we test how much identification potential they have on their own. %

For \textbf{E3} we use the SMPL body model fits from both VPoser and HuMoR for our motion data where specific components are removed by setting them to zero. We then generate new position data for the classification.
We expect the body shape and poses alone to carry a high identification potential, as has been shown in previous studies~\cite{DBLP:journals/popets/HanischMHLS23, sattar2020body}. For the root translation and especially for the root rotation we expect a lower identification potential, because these are single vectors, which should carry less information than the body poses.

\paragraph{Dependency Reduction}
It is our assumption that by encoding our motion sequences using the SMPL body model and then foundation motion models, the dependencies between the individual data dimensions of a pose or pose transition are reduced due to the \ac{VAE} architecture used in the foundation models.
For \textbf{E4} we measure the linear dependence of the pose dimensions on each other. We do this by measuring the average absolute covariance between all dimensions of a pose and then averaging them over the number of poses per motion sequence. This gives us a single comparable linear dependency measure per motion sequence. We then compare the dependency for the different encodings of the motion sequence (original and latent code) to see if the dependency is reduced by the encoding of the foundation motion models.

\paragraph{Noise Mode Comparison}
We assume that adding the same noise vector to all poses of a motion sequence (static noise mode) will perform better than adding a new noise vector to each pose (variable noise mode). In our noise mode comparison experiment \textbf{E5}, we run both modes with different values of the noise scaling $\gamma$ on the original motion data to test this assumption. We then measure privacy by performing identification with our biometric recognition system. We expect that the static noise mode will always outperform the variable noise mode. Since a motion sequence is a time series of poses, two consecutive poses will be very similar to each other because a person cannot move much in a single time step. Adding different noise vectors to two similar poses makes it easier to separate the noise from the underlying data, since much of the difference between the two poses after the noise is added is the noise. If we add the same noise vector to both poses, then the difference between the two poses is still the same as it was before anonymization, and we cannot distinguish noise from data.

\subsubsection{Privacy-Utility Evaluation}

In our privacy-utility experiment \textbf{E6} we evaluate the privacy and utility of Pantomime. For a better comparison, we not only study how noise injection into the latent space affects the privacy-utility tradeoff, but also test applying noise directly to the original data and the fitted SMPL representation. Due to the different representation of the motion sequence (original, SMPL fit, latent code), the noise parameters of our different anonymization techniques are not directly comparable.
For example, adding the same amount of noise to the position of a joint will have a different effect than adding noise to the joint rotations of an SMPL fit.
In order to achieve comparability between where we apply noise, we define protection targets.
A protection target is a given value of recognition accuracy, for example 20\%.
We then tune the noise parameters of our anonymization to achieve the given target ($\pm 2\%$).
In this way, the anonymization performance of our techniques is the same, and we can directly compare the utility of our approaches to judge which anonymization has the better tradeoff.
An overview of all combinations of motion representation, anonymization, and protection target is given in Table~\ref{tab:overview_experiments}.

We investigate utility by performing action recognition on the CeTI-Locomotion dataset and by conducting a user study on both datasets. We investigate two utility goals in our user study, the first is the naturalness of an anonymized motion sequence, and the second is the motion similarity between an original and corresponding anonymized motion sequence. %

\begin{table}
\centering
\caption{The combinations of motion representations, anonymizations, and protection targets which we use in our privacy-utility tradeoff.}\label{tab:overview_experiments}
\begin{tabular}{|l|l|l|} 
\hline
motion rep. & anonymization  & protection targets  \\ 
\hline
original & direct & 10\%, 20\%           \\ 
\hline
SMPL (VPoser) & direct, VPoser, HuMoR & 10\%, 20\%           \\ 
\hline
SMPL (HuMoR) & direct, VPoser, HuMoR & 10\%, 20\%           \\
\hline
\end{tabular}
\end{table}

VPoser only works on a single pose at a time, while HuMoR works on the pose transition and thus on pose pairs. Due to this, and the better overall performance for pose fitting reported by HuMoR~\cite{rempe2021humor}, we expect HuMoR to do a better job of removing the dependencies, and therefore Pantomime to achieve better utility. Furthermore, we expect Pantomime to achieve better utility than adding noise to the original pose data, regardless of the foundation motion model chosen.

\subsection{Utility / User Study} %

\begin{figure}[t]
    \centering
    \includegraphics[width=\linewidth]{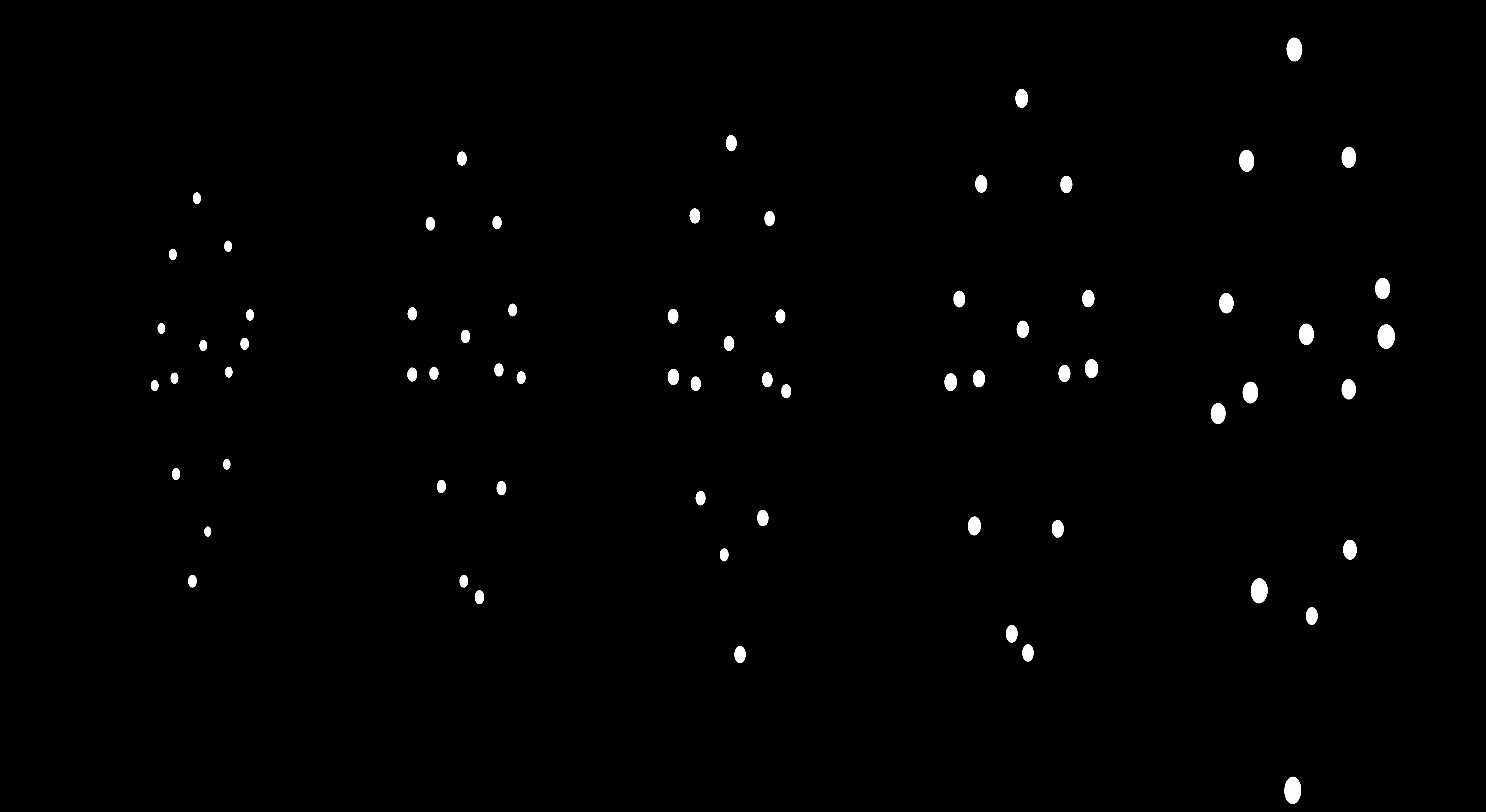}
   \caption{Example rendering of a motion sequence used during the user study. Link to all animated videos during the user study: \url{http://49.13.59.7/} 
   }\label{fig:user-study-comp}
\end{figure}

For utility, we investigate how natural the anonymized motion sequences appear and how similar the actions are compared to the original motion sequence. We do this by conducting a user study similar to prior work~\cite{DBLP:journals/popets/TodtHS24} on the evaluation of biometric anonymizations. We use two tasks to evaluate our objectives. In the first task, we show participants a single motion sequence. The participants then rate how natural the sequences appear by answering the question "Is this a natural human motion? The rating is done on a 5-point Likert scale from "very unnatural" (1) to "very natural" (5). In the second task, we show the original sequence next to an anonymized version of the sequence. Participants then answer the question "How similar are the motions performed in the two videos?" by rating on a 5-point Likert scale from "very dissimilar" (1) to "very similar" (5).

We conducted an online survey in which participants were shown different motion sequences from the CeTI-Locomotion and Horst-DB datasets as short video sequences (see Figure~\ref{fig:user-study-comp} for an example) rendered at 20 frames per second. We reduced the rendering from 30 to 20 frames per second to allow users to better judge the execution of the motion. The motion is rendered as point-light displays to reduce the influence of appearance and to focus on the motion. From the Horst DB, we randomly select 4 motion sequences, two from male participants and two from female participants. From CeTI-Locomotion, we randomly select two male and two female motion sequences for the modalities \textit{gait-normal}, \textit{gait-fast}, \textit{gait-bottle-crate}, and \textit{sit-to-stand}. These 20 motion sequences were anonymized with a subset (original+direct, HuMoR+direct, HuMoR+HuMoR, VPoser+direct, and VPoser+VPoser) of the combinations described above (see Table~\ref{tab:overview_experiments}) for the protection targets of 10\% and 20\%. We publish an overview\footnote{\url{http://49.13.59.7/}} of the motion sequences used and their anonymized variants alongside this paper. Since we perform two tasks (naturalness and similarity see above) and add the original sequences, this results in 440 unique questions. From this question pool, each participant answers 40 random questions.

\subsection{Ethical Considerations}

The user study data collection was approved by the ethics commission of \ifanon Anon University \else the Karlsruhe Institute of Technology (research project "Utility of Anonymized Motion Sequences") \fi and was conducted in accordance with the Declaration of Helsinki. All data was collected in an anonymous online survey in December 2024 using an online recruitment platform\footnote{https://prolific.com} to recruit 224 participants (112 male, 112 female; mean age 30.8, std 9.57). Participation took a median of 7:47 minutes and participants were paid an average of 10.71£ per hour.

The CeTI-Locomotion and Horst-DB datasets used in this study were both approved by their respective ethics committees and their participants gave written informed consent to participate in the data collection.

\section{Results}\label{sec:results}

Here, we describe the results of our experiments, similar to the evaluation section, where we start with the results of testing the underlying assumptions for Pantomime before describing the results of the privacy-utility evaluation.

\subsection{Assumptions \& Baseline}

We start with the baseline identification experiment \textbf{E1} and the action recognition experiment \textbf{E2}. For both CeTI-Locomotion and Horst-DB we observe that the identification recognition accuracy for the original position data is high with 83\% and 96\%, respectively. For the CeTI-Locomotion dataset, we also perform action recognition to classify which of the 5 actions was performed in a motion sequence, achieving an accuracy of 80\%. Overall, these results are in line with prior work and show that the recognition systems used for identification and action recognition work.

\providecommand{\vposertranscetionly}{0.65}
\providecommand{\vposertranscetiwithout}{0.39}
\providecommand{\vposertranshorstonly}{1.0}
\providecommand{\vposertranshorstwithout}{0.51}
\providecommand{\vposershapecetionly}{0.52}
\providecommand{\vposershapecetiwithout}{0.59}
\providecommand{\vposershapehorstonly}{0.75}
\providecommand{\vposershapehorstwithout}{0.5}
\providecommand{\vposerorientcetionly}{0.61}
\providecommand{\vposerorientcetiwithout}{0.19}
\providecommand{\vposerorienthorstonly}{0.75}
\providecommand{\vposerorienthorstwithout}{0.93}
\providecommand{\vposerposecetionly}{0.63}
\providecommand{\vposerposecetiwithout}{0.26}
\providecommand{\vposerposehorstonly}{0.69}
\providecommand{\vposerposehorstwithout}{1.0}
\providecommand{\humortranscetionly}{0.21}
\providecommand{\humortranscetiwithout}{0.2}
\providecommand{\humortranshorstonly}{0.64}
\providecommand{\humortranshorstwithout}{0.39}
\providecommand{\humorshapecetionly}{0.38}
\providecommand{\humorshapecetiwithout}{0.1}
\providecommand{\humorshapehorstonly}{0.6}
\providecommand{\humorshapehorstwithout}{0.11}
\providecommand{\humororientcetionly}{0.31}
\providecommand{\humororientcetiwithout}{0.06}
\providecommand{\humororienthorstonly}{0.55}
\providecommand{\humororienthorstwithout}{0.19}
\providecommand{\humorposecetionly}{0.2}
\providecommand{\humorposecetiwithout}{0.18}
\providecommand{\humorposehorstonly}{0.41}
\providecommand{\humorposehorstwithout}{0.47}

\definecolor{vposertranscetionly}{rgb}{0.185783,0.704891,0.485273}
\definecolor{vposertranscetiwithout}{rgb}{0.168126,0.459988,0.558082}
\definecolor{vposertranshorstonly}{rgb}{0.993248,0.906157,0.143936}
\definecolor{vposertranshorstwithout}{rgb}{0.125394,0.574318,0.549086}
\definecolor{vposershapecetionly}{rgb}{0.122606,0.585371,0.546557}
\definecolor{vposershapecetiwithout}{rgb}{0.130067,0.651384,0.521608}
\definecolor{vposershapehorstonly}{rgb}{0.369214,0.788888,0.382914}
\definecolor{vposershapehorstwithout}{rgb}{0.127568,0.566949,0.550556}
\definecolor{vposerorientcetionly}{rgb}{0.143303,0.669459,0.511215}
\definecolor{vposerorientcetiwithout}{rgb}{0.258965,0.251537,0.524736}
\definecolor{vposerorienthorstonly}{rgb}{0.369214,0.788888,0.382914}
\definecolor{vposerorienthorstwithout}{rgb}{0.82494,0.88472,0.106217}
\definecolor{vposerposecetionly}{rgb}{0.162016,0.687316,0.499129}
\definecolor{vposerposecetiwithout}{rgb}{0.225863,0.330805,0.547314}
\definecolor{vposerposehorstonly}{rgb}{0.24607,0.73891,0.452024}
\definecolor{vposerposehorstwithout}{rgb}{0.993248,0.906157,0.143936}
\definecolor{humortranscetionly}{rgb}{0.250425,0.27429,0.533103}
\definecolor{humortranscetiwithout}{rgb}{0.253935,0.265254,0.529983}
\definecolor{humortranshorstonly}{rgb}{0.170948,0.694384,0.493803}
\definecolor{humortranshorstwithout}{rgb}{0.168126,0.459988,0.558082}
\definecolor{humorshapecetionly}{rgb}{0.171176,0.45253,0.557965}
\definecolor{humorshapecetiwithout}{rgb}{0.282623,0.140926,0.457517}
\definecolor{humorshapehorstonly}{rgb}{0.134692,0.658636,0.517649}
\definecolor{humorshapehorstwithout}{rgb}{0.281412,0.155834,0.469201}
\definecolor{humororientcetionly}{rgb}{0.201239,0.38367,0.554294}
\definecolor{humororientcetiwithout}{rgb}{0.281924,0.089666,0.412415}
\definecolor{humororienthorstonly}{rgb}{0.119423,0.611141,0.538982}
\definecolor{humororienthorstwithout}{rgb}{0.258965,0.251537,0.524736}
\definecolor{humorposecetionly}{rgb}{0.253935,0.265254,0.529983}
\definecolor{humorposecetiwithout}{rgb}{0.262138,0.242286,0.520837}
\definecolor{humorposehorstonly}{rgb}{0.160665,0.47854,0.558115}
\definecolor{humorposehorstwithout}{rgb}{0.13777,0.537492,0.554906}

\begin{table}
\centering
\small
\caption{The influence of the individual parameters of the VPoser and Humor SMPL fits on the identification given as accuracy}\label{tab:component_analysis}
\begin{tblr}{
  cell{1}{1} = {c=2}{},
  cell{2}{1} = {r=4}{},
  cell{2}{3} = {vposershapecetionly, fg=white},
  cell{2}{4} = {vposershapehorstonly, fg=white},
  cell{3}{3} = {vposerposecetionly, fg=white},
  cell{3}{4} = {vposerposehorstonly, fg=white},
  cell{4}{3} = {vposertranscetionly, fg=white},
  cell{4}{4} = {vposertranshorstonly, fg=black},
  cell{5}{3} = {vposerorientcetionly, fg=white},
  cell{5}{4} = {vposerorienthorstonly, fg=white},
  cell{6}{1} = {r=4}{},
  cell{6}{3} = {humorshapecetionly, fg=white},
  cell{6}{4} = {humorshapehorstonly, fg=white},
  cell{7}{3} = {humorposecetionly, fg=white},
  cell{7}{4} = {humorposehorstonly, fg=white},
  cell{8}{3} = {humortranscetionly, fg=white},
  cell{8}{4} = {humortranshorstonly, fg=white},
  cell{9}{3} = {humororientcetionly, fg=white},
  cell{9}{4} = {humororienthorstonly, fg=white},
  vlines,
  hline{1-2,6,10} = {-}{},
  hline{3-5,7-9} = {2-4}{},
}

\diagbox{SMPL para.}{datasets} &              & CeTI-Locomotion & Horst-DB \\
VPoser     & shape~($\beta$) & \vposershapecetionly & \vposershapehorstonly \\
           & joint poses~($\Theta$)  & \vposerposecetionly & \vposerposehorstonly \\
           & root trans.~($r$)  & \vposertranscetionly & \vposertranshorstonly \\
           & root orient.~($\Phi$) & \vposerorientcetionly & \vposerorienthorstonly \\
HuMoR      & shape~($\beta$)        & \humorshapecetionly & \humorshapehorstonly \\
           & joint poses~($\Theta$)  & \humorposecetionly & \humorposehorstonly \\
           & root trans.~($r$)  & \humortranscetionly & \humortranshorstonly \\
           & root orient.~($\Phi$) & \humororientcetionly & \humororienthorstonly \\  
\end{tblr}
\end{table}

For \textbf{E3} we look at the difference in identification accuracy when using only certain parts of the data in their SMPL representation. In Table~\ref{tab:component_analysis} we report the identification accuracy for the data when using positions generated from the SMPL representation with only the specific parameter intact, while the rest of the parameters are set to zero. We find that most parameters have a high identification potential (greater than 50\%) on their own. The lowest identification potential is observed for the SMPL fit of the CeTI-Locomotion data using HuMoR. But even here, the individual parameters have significantly more identification potential than the chance level (2\%) for CeTI-Locomotion. This leads us to conclude that anonymizing only the SMPL joint poses in the latent space is not sufficient, as the remaining components of the SMPL body model can be used for identification. This justifies our decision to set the remaining parameters to zero.

\providecommand{\corcetihumorpositions}{0.62}
\providecommand{\corcetihumorposes}{0.55}
\providecommand{\corcetihumorvposer}{0.52}
\providecommand{\corcetihumorhumor}{0.37}
\providecommand{\corcetivposerpositions}{0.55}
\providecommand{\corcetivposerposes}{0.55}
\providecommand{\corcetivposervposer}{0.5}
\providecommand{\corcetivposerhumor}{0.4}

\definecolor{corcetihumorpositions}{rgb}{0.150148,0.676631,0.506589}
\definecolor{corcetihumorposes}{rgb}{0.119423,0.611141,0.538982}
\definecolor{corcetihumorvposer}{rgb}{0.122606,0.585371,0.546557}
\definecolor{corcetihumorhumor}{rgb}{0.175841,0.44129,0.557685}
\definecolor{corcetivposerpositions}{rgb}{0.119423,0.611141,0.538982}
\definecolor{corcetivposerposes}{rgb}{0.119423,0.611141,0.538982}
\definecolor{corcetivposervposer}{rgb}{0.127568,0.566949,0.550556}
\definecolor{corcetivposerhumor}{rgb}{0.163625,0.471133,0.558148}

\providecommand{\corhorsthumorpositions}{0.64}
\providecommand{\corhorsthumorposes}{0.5}
\providecommand{\corhorsthumorvposer}{0.46}
\providecommand{\corhorsthumorhumor}{0.39}
\providecommand{\corhorstvposerpositions}{0.46}
\providecommand{\corhorstvposerposes}{0.55}
\providecommand{\corhorstvposervposer}{0.41}
\providecommand{\corhorstvposerhumor}{0.34}

\definecolor{corhorsthumorpositions}{rgb}{0.170948,0.694384,0.493803}
\definecolor{corhorsthumorposes}{rgb}{0.127568,0.566949,0.550556}
\definecolor{corhorsthumorvposer}{rgb}{0.141935,0.526453,0.555991}
\definecolor{corhorsthumorhumor}{rgb}{0.168126,0.459988,0.558082}
\definecolor{corhorstvposerpositions}{rgb}{0.141935,0.526453,0.555991}
\definecolor{corhorstvposerposes}{rgb}{0.119423,0.611141,0.538982}
\definecolor{corhorstvposervposer}{rgb}{0.160665,0.47854,0.558115}
\definecolor{corhorstvposerhumor}{rgb}{0.187231,0.414746,0.556547}

\begin{table}
\centering
\small
\caption{Average absolute linear correlation between all of the dimensions of the respective encoding of a motion sequence.}\label{tab:dependency_analysis}
\begin{tblr}{
  cell{1}{2} = {c=2}{},
  cell{1}{4} = {c=2}{},
  cell{3}{2} = {corcetivposerpositions, fg=white},
  cell{3}{3} = {corcetihumorpositions, fg=white},
  cell{3}{4} = {corhorstvposerpositions, fg=white},
  cell{3}{5} = {corhorsthumorpositions, fg=white},
  cell{4}{2} = {corcetivposerposes, fg=white},
  cell{4}{3} = {corcetihumorposes, fg=white},
  cell{4}{4} = {corhorstvposerposes, fg=white},
  cell{4}{5} = {corhorsthumorposes, fg=white},
  cell{5}{2} = {corcetivposervposer, fg=white},
  cell{5}{3} = {corcetihumorvposer, fg=white},
  cell{5}{4} = {corhorstvposervposer, fg=white},
  cell{5}{5} = {corhorsthumorvposer, fg=white},
  cell{6}{2} = {corcetivposerhumor, fg=white},
  cell{6}{3} = {corcetihumorhumor, fg=white},
  cell{6}{4} = {corhorstvposerhumor, fg=white},
  cell{6}{5} = {corhorsthumorhumor, fg=white},
  hlines,
  vlines,
}
                 & CeTI-Locomotion &       & Horst-DB &       \\
SMPL Fit         & VPoser          & HuMoR & VPoser   & HuMoR \\
positions        & \corcetivposerpositions   & \corcetihumorpositions  & \corhorstvposerpositions     & \corhorsthumorpositions  \\
joint poses      & \corcetivposerposes & \corcetihumorposes  & \corhorstvposerposes     & \corhorsthumorposes  \\
VPoser lat. enc. & \corcetivposervposer & \corcetihumorvposer  & \corhorstvposervposer     & \corhorsthumorvposer  \\
HuMoR lat enc.   & \corcetihumorvposer & \corcetihumorhumor  & \corhorstvposerhumor      & \corhorsthumorhumor  
\end{tblr}
\end{table}

For \textbf{E4} we hypothesized that using the foundation motion models will reduce the dependency between the individual data dimensions. In Table~\ref{tab:dependency_analysis} we report the average absolute correlation between the different data dimensions for different representations of the motion sequences. We find that the data represented as positions (generated from the respective SMPL fit) has the highest average absolute correlation, with the exception of the Horst-DB VPoser fit. The data represented as joint poses has the second highest, followed by the latent encodings of VPoser and HuMoR. This decrease in correlation is in line with our expectations, and especially HuMoR seems to achieve a high decoupling of the latent dimensions. However, the remaining high correlations for the VPoser latent encoding show that the effects of decoupling can be much smaller than expected.

\begin{figure}[!tb]
    \centering
    \includegraphics[width=\linewidth]{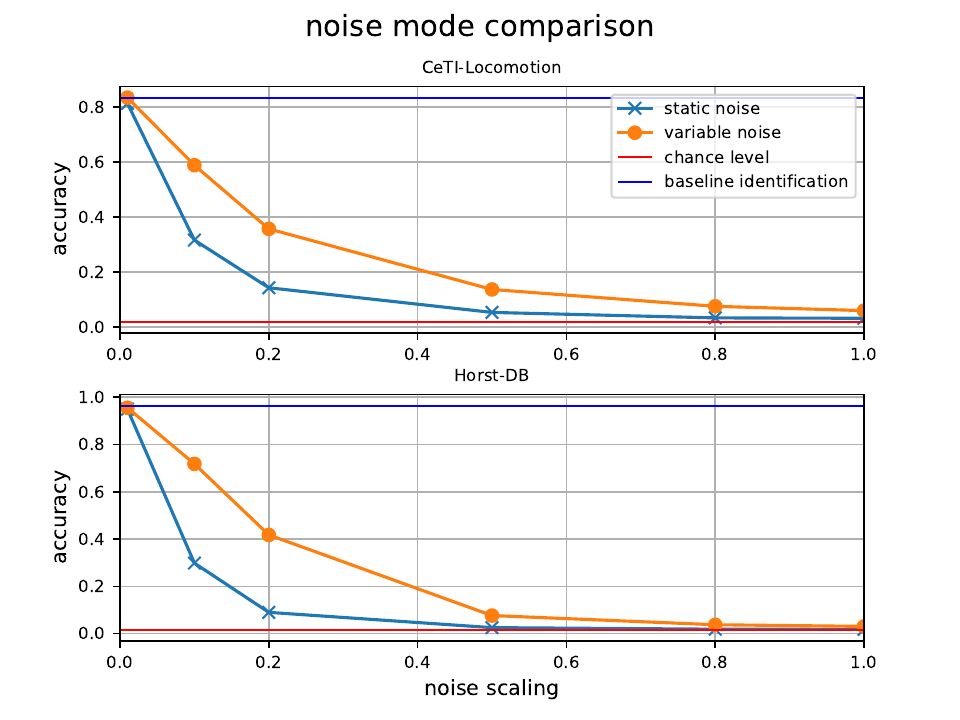}
    \caption{Comparison of the static noise vs. variable noise mode of directly applying noise to the original data of both CeTI-Locomotion and Horst-DB over different $\gamma$ noise scaling choices.}\label{fig:modus-all}
\end{figure}

Next, we report the results for our noise mode experiment \textbf{E5}, see Figure~\ref{fig:modus-all}. For both datasets, we can see that the variable noise mode is always outperformed by the static noise mode when using the same noise scaling $\gamma$. When visually inspecting the resulting motion sequences using a rendering of the sequence, it can also be seen that the variable approach leads to a very visible shaking of the joint points, the injected noise becomes visible in the motion execution.

Since two consecutive poses are very similar to each other, changing the noise vector for each pose makes it easier to distinguish what is the real data and what is the noise addition. Our conclusion from this experiment is that the static noise mode is the better mode to add noise to the data because it always outperforms the variable noise for identification reduction.

\subsection{Privacy-Utility Results}

\begin{figure}[!bt]
    \includegraphics[width=\linewidth]{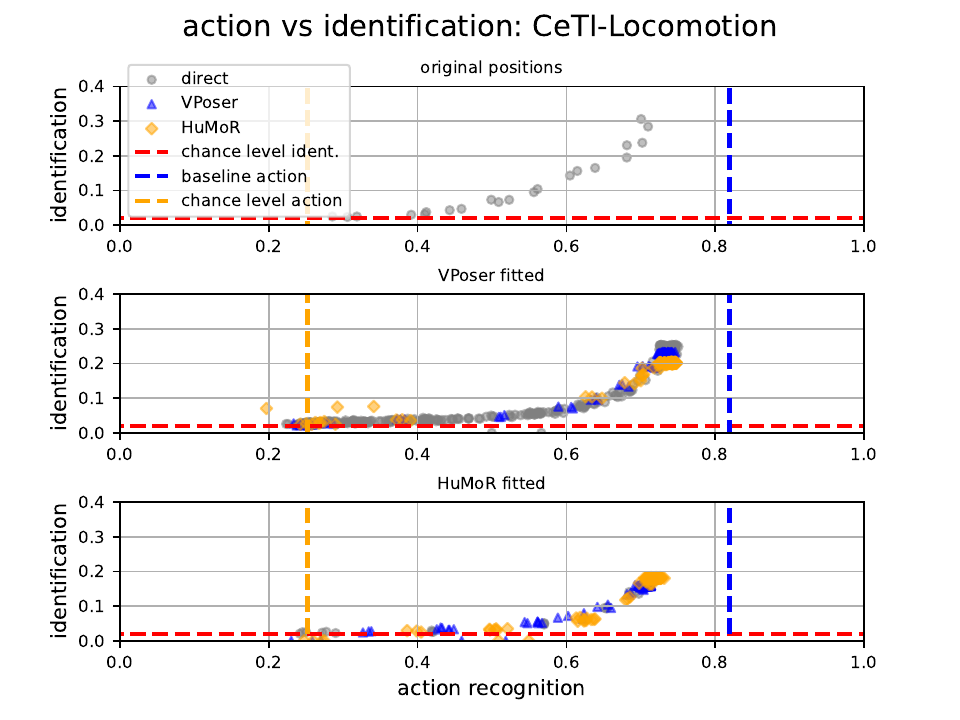}
    \caption{The action recognition vs identity recognition accuracy for CeTI-Locomotion.}\label{fig:action_vs_identification_accuracy}
    \label{fig:action_vs_recognition}
\end{figure}

To investigate the privacy-utility tradeoff \textbf{E6}, we first report the action recognition results for the CeTI-Locomotion dataset, see Figure~\ref{fig:action_vs_identification_accuracy}, before doing the main comparison using our user study. Comparing the direct anonymization on the original data with all the anonymizations on the adjusted data, we see that the adjusted data drops faster in person identification than in action recognition. For example, at 60\% action recognition accuracy, the original data is still at about 15\% identification accuracy, while the fitted data is much lower and close to 5\% accuracy. Surprisingly, there does not seem to be a difference if the anonymization on the fitted data is performed directly on the positions, the SMPL joints, or the latent encoding, as all these anonymizations perform similarly. At least for action recognition, it does not seem to matter in which data space the anonymization is performed. We conclude that action recognition is a simple task that is still successful on heavily distorted data, which leads to similar performance of the anonymizations.

\begin{figure}[!tb]
    \includegraphics[width=\linewidth]{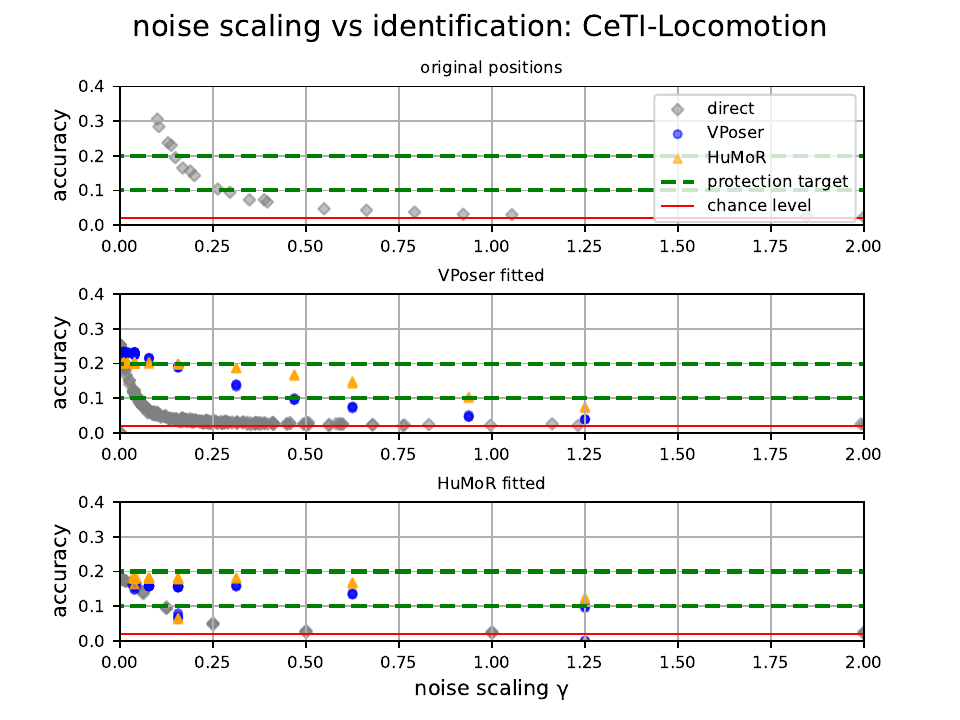}
    \caption{The scaling of the noise parameter $\gamma$ in comparison to the achieved recognition accuracy of different anonymization combinations for CeTI-Locomotion.}\label{fig:noise_rec_ceti}
\end{figure}

\begin{figure}[!tb]
    \includegraphics[width=\linewidth]{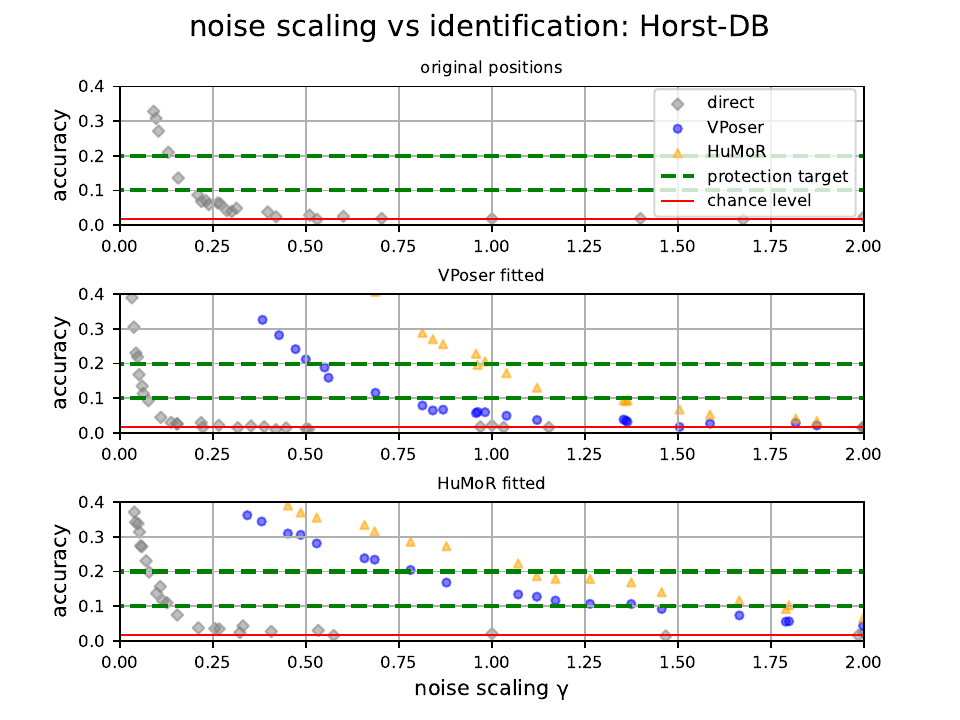}
    \caption{The scaling of the noise parameter $\gamma$ in comparison to the achieved recognition accuracy of different anonymization combinations for Horst-DB.}\label{fig:noise_rec_horst}
\end{figure}

For the main privacy-utility tradeoff, we first report the scaling of the noise parameter versus the achieved recognition accuracy, see Figure~\ref{fig:noise_rec_ceti} and Figure~\ref{fig:noise_rec_horst}. We use these results to pick the noise scaling values which fulfill the specific protection targets (shown as green line). We find that the direct application of noise to the position data requires less noise scaling than when we apply the noise in the latent space or directly to the joint rotations of the SMPL model. For both datasets, we choose the anonymizations that achieve 10\% or 20\% ($\pm2\%$) for the direct comparison of utility in the user study. %

\begin{figure}[!tb]
    \centering
    \includegraphics[width=\linewidth]{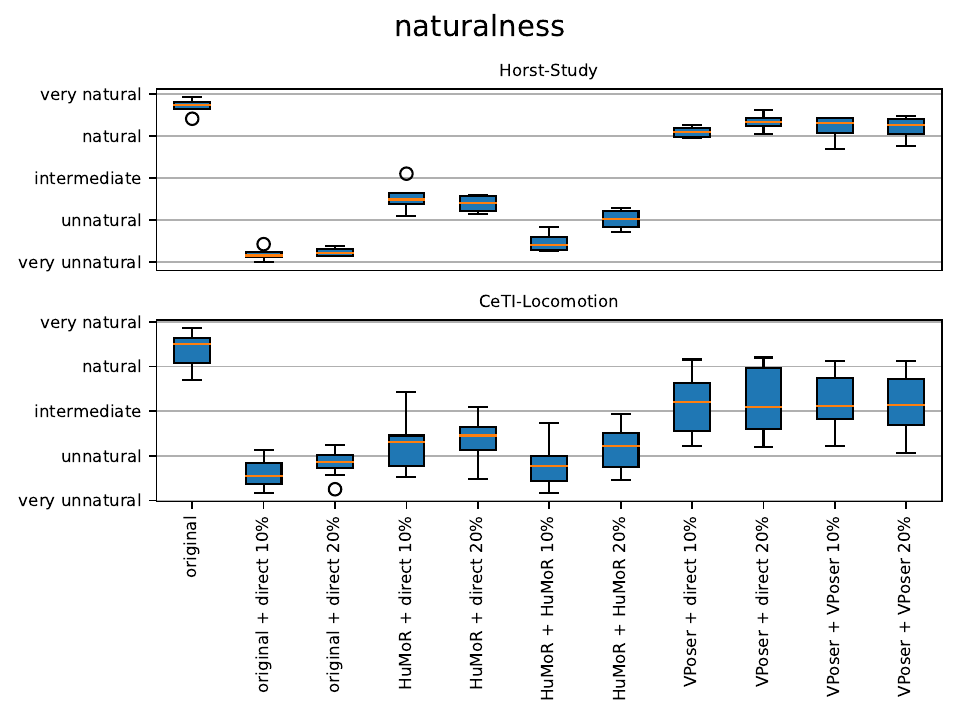}
    \caption{Average naturalness rating per question of the user study as a box plot per anonymization.}\label{fig:naturalness-user-study}
\end{figure}

\begin{figure}[!tb]
    \centering
    \includegraphics[width=\linewidth]{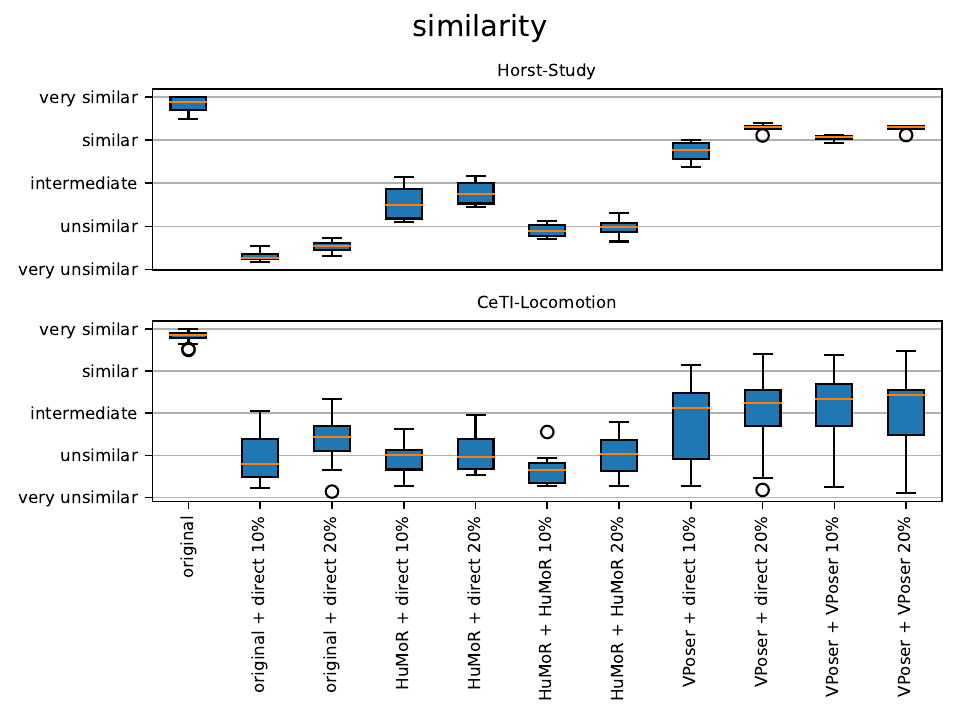}
    \caption{Average similarity rating per question of the user study as a box plot per anonymization.}\label{fig:similarity-user-study}
\end{figure}

For our user study, we first calculate the average of the ratings for each question, and then use the averages to create a box plot for each anonymization technique. Comparing the naturalness results (see Figure~\ref{fig:naturalness-user-study}) we see that the original motion sequences score the highest for both datasets, indicating that the participants consider the selected datasets to be representative of natural motion. For the direct addition of noise to the original position data, we see that both datasets have very low naturalness ratings, close to very unnatural, showing that the naturalness of the motion data is destroyed. The anonymizations using HuMoR as a foundation model achieve more utility than the direct anonymization, but also all are rated close to unnatural. The best natural ratings we see for the anonymization using VPoser, for the CeTI-Locomotion dataset an intermediate rating is achieved, while for the Horst-DB the rating is even above natural. We conclude that using VPoser as a foundation model during anonymization helps to achieve natural motion sequences that are close to the naturalness of the original.

For the similarity results (see Figure~\ref{fig:similarity-user-study}), we generally see a similar pattern as for naturalness, with the original data rated as very similar to itself. Direct anonymization on the original data achieves almost no similarity, HuMoR achieves intermediate to unsimilar results, and VPoser again delivers the best similarity on the Horst-DB datasets. This again shows that using VPoser, Pantomime can successfully anonymize while keeping the motion sequence similar to the original, thus preserving utility.

\subsection{Summary of Results}\label{sec:summary_of_results}%

\begin{itemize}
    \item Pantomime is able to successfully anonymize motion data by anonymizing it in the latent space of a foundation motion model.
    \item All components of the SMPL representation of the motion sequences contain identifiable information.
    \item The latent encodings of motion sequences using foundation motion models only slightly reduce the correlations between the data dimensions.
    \item Applying a fixed random vector to the entire motion sequence instead of varying it for each pose is the better mode for anonymizing motion sequences.
    \item For the action recognition we do not see a significant difference between the anonymization techniques.
    \item Using a VPoser fitting with a VPoser latent encoding achieves the best privacy-utility tradeoff.
\end{itemize}

\section{Discussion, Limitations \& Future Work}\label{sec:discussion}

In general, we find that Pantomime's approach of using foundation motion models to first fit the position data to the SMPL body model and then anonymize the data in the latent space of the model is a viable approach to anonymize full-body motion data in a plausible manner. Because of the plausibility constraints added by the foundation models, the data can retain the utility of the motion data while performing effective anonymization.

In comparison with prior work, Pantomime has some key advantages. It anonymizes full-body motion capture data, it is not designed to work only with specific motions (such as gait), it does not require a large corpus of specific application data to train, and it is configurable via its noise scaling, allowing its privacy-utility tradeoff to be adjusted for specific applications. Furthermore, it is general as it performs a unification of motion data formats by fitting to the SMPL body model.

The same foundation models that we use to anonymize can also be used to generate synthetic motion data, by using the original data as an anchor in the latent space and then shifting it by adding noise, we essentially generate synthetic data that is similar to the original data. Thus, Pantomime can be considered as synthetic data generation. Similar to face anonymization~\cite{hukkelaas2019deepprivacy}, which generates new faces to anonymize facial images that have similar characteristics such as ethnicity or age.

Pantomime also has some limitations that need to be addressed. Its main drawback is the poor fitting quality for some of the motion sequences, especially when HuMoR is used for fitting. Furthermore, the fitting process is slow in its current implementation (about 1.5 weeks for CeTI-Locomotion on a single GeForce RTX 3090). Another problem with fitting is that it does not work well for certain actions, such as the stand-to-sit tasks in CeTI-Locomotion.

Pantomime's approach to anonymization is very general and could be promising for other complex data types that are difficult to anonymize, such as trajectories or other biometric features. The only requirement would be that the foundation models use a \ac{VAE} structure. Furthermore, it should be investigated how the approach can be realized in a faster fashion to allow for applications like the real-time streaming of motion data.

\section{Conclusion}\label{sec:conclusion}

We have presented Pantomime, a full-body motion anonymization that uses foundation motion models to anonymize in a plausible way. Our results show that using the VPoser model it is possible to achieve identification rates as low as 10\% while keeping the anonymized motions natural and similar to the original ones. This is an important step towards a more privacy-preserving use of motion tracking in applications such as \ac{MR}, robotics, or medicine.

\ifCLASSOPTIONcompsoc
  \section*{Acknowledgments}
  \ifanon
  \else
  Funded by the German Research Foundation (DFG, Deutsche Forschungsgemeinschaft) as part of Germany’s Excellence Strategy – EXC 2050/1 – Project ID 390696704 – Cluster of Excellence “Centre for Tactile Internet with Human-in-the-Loop” (CeTI) of Technische Universität Dresden; This work was funded by the Topic Engineering Secure Systems of the Helmholtz Association (HGF) and supported by KASTEL Security Research Labs, Karlsruhe.
  \fi
\else
  \section*{Acknowledgment}
\fi

\bibliographystyle{ieeetr}
\bibliography{pantomime}

\end{document}